\documentclass[prb,aps,amsmath,amssymb,showpacs,twocolumn]{revtex4}
\usepackage{graphicx}
\usepackage{booktabs}

\newcommand{\mean}[1]{{\langle{#1}\rangle}}
\newcommand{\cmref}[1]{\href{http://arXiv.org/abs/cond-mat/#1}{cond-mat/#1}} 

\begin{document}

\title{Thermodynamic properties and correlation functions of Ar films on the surface of a bundle of nanotubes}
\author{Nathan M. Urban}
\email{nurban@phys.psu.edu}
\author{Silvina M. Gatica}
\email{gatica@phys.psu.edu}
\author{Milton W. Cole}
\email{mwc@psu.edu}
\affiliation{Department of Physics, Pennsylvania State University, University Park, Pennsylvania 16802-6300, USA}
\author{Jos\'e L. Riccardo}
\email{jlr@unsl.edu.ar}
\affiliation{Departamento de F{\'\i}sica, Universidad Nacional de San Luis, 5700 San Luis, Argentina}

\begin{abstract}
We employ canonical Monte Carlo simulations to explore the properties
of an Ar film adsorbed on the external surface of a bundle of carbon
nanotubes.  The study is concerned primarily with three properties:
specific heat $c(T)$, differential heat of adsorption $q_d$, and
Ar-Ar correlation functions $g(\mathbf{r})$.  These measurable
functions exhibit information about the dependence of film structure
on coverage and temperature.
\end{abstract}


\maketitle

\section{Introduction}
\label{sec:intro}

Considerable interest has been attracted recently to the properties
of simple gases (noble gases and small molecules) adsorbed near
bundles of carbon
nanotubes.\cite{byl-etal,bienfait-etal,bienfait-etal2,jakubek-simard,talapatra-krungleviciute-migone,talapatra-migone,talapatra-rawat-migone,pradhan-etal,ramachandran-etal,kahng-etal,wilson-etal,wilson-thesis,gatica-etal1,migone,smith-russo-etal,smith-bittner-etal,lasjaunias-etal,siber1}
This subject has been reviewed recently.\cite{gatica-etal1,migone}
The adsorption of these gases can occur within the tubes only if
they are open, which is possible either during the process of
nanotube formation (e.g., when endohedral C$_{60}$ molecules are
formed),~\cite{smith-russo-etal} or after chemical treatment to open
the tubes.\cite{smith-bittner-etal,kuznetsova-etal}  The presence,
or absence, of interstitial channel (IC) molecules is an open
question in the case of an idealized bundle of identical tubes;
there seems to be no doubt, however, that such IC adsorption occurs
in laboratory samples of polydisperse tubes.\cite{polydisperse}
In contrast to variability in adsorption at these sites, the
adsorption of gas on the external surface of the bundle is a
ubiquitous phenomenon, in which the film coverage increases with
the pressure ($P$) of the coexisting gas.  In that exohedral
environment, an adatom is strongly attracted to the groove region
between two neighboring tubes; there, the film forms a
quasi-one-dimensional phase.  Further adsorption at low temperature
($T$) is predicted to manifest a so-called three-stripe phase of gas
aligned parallel to the grooves.\cite{three-stripe}  At higher gas
coverage ($N$), there occurs a two-dimensional monolayer phase,
qualitatively analogous to that found on the graphite
surface.\cite{dash-etal,phys-ad-book}  At even higher coverage, a
multilayer film grows as $P$ increases.  There is an upper limit
of total film coverage, set by the bundle's
curvature;\cite{taborek,hernandez-etal} this limit has yet to be
explored.

This study extends a previous investigation~\cite{gatica-etal2} of
the adsorption of Ar gas on the external surface of a nanotube
bundle.  Argon was chosen as a model adsorbate because its gas-gas
interaction is well known, making it a standard fluid in the study
of simple fluids.  In the previous study, denoted I, we employed
the grand canonical Monte Carlo simulation method to explore the
evolution of the equilibrium film as a function of $P$ and $T$.
The present paper, stimulated by recent and proposed experiments,
adds three results to those derived in the previous study.  One
property is the specific heat, $c(T)$, which is computed here from
energy fluctuations, evaluated using simulations within the canonical
ensemble.  The second property is the differential heat of adsorption,
$q_d = -(\partial{E}/\partial{N})_T$, where $E$ is the energy of
the film.  This quantity is closely related to another quantity,
which is more often measured experimentally, the isosteric heat
$q_{st}=(\partial{(\ln{P})}/\partial{\beta})_N$ [where $\beta =
1/(k_B T)$], by the relation~\cite{elgin-goodstein} $q_{st} = q_d
+ k_B T$ (assuming an ideal gas coexisting with the film).  The
third property reported here is the anisotropic correlation function
of the overlayer.  This quantity is related by Fourier transform
to results of diffraction experiments.  With the exception of the
isosteric heat calculated by Shi and Johnson,\cite{polydisperse}
none of these properties has been explored in simulation studies
of films on nanotube bundles, prior to the present work.

The outline of this paper is the following.  Section~\ref{sec:comp-methods}
summarizes our simulation methods.  Section~\ref{sec:dens-corr}
reports results for the density and correlation functions.
Section~\ref{sec:heat-cap} presents results for the thermodynamic
properties, $c$ and $q_d$.  Section~\ref{sec:summary} summarizes
our results.

\section{Computational methods}
\label{sec:comp-methods}

When not explicitly contradicted in this paper, it may be assumed
that the physical system and computational method are as described
in I.  The primary model system is a bundle of infinitely long,
cylindrically symmetric carbon nanotubes of identical radii equal
to 6.9~{\AA}.  Only two adjacent nanotubes on the external surface
of the bundle are simulated.  The $y$ axis is parallel to the
nanotubes, and the $z$ axis is directed away from the surface of
the bundle.  Periodic boundary conditions are imposed in the $x$
and $y$ directions (approximating the surface of the bundle as an
infinite plane of nanotubes); reflecting boundary conditions are
imposed in the $z$ direction.  The unit simulation cell, whose
volume contains half of each of the two adjacent nanotubes with the
groove in between them at the center, is 17~{\AA} in the $x$
direction, 34~{\AA} in the $y$ direction, and 40~{\AA} in the $z$
direction.

The simulations were done in the canonical ensemble, rather than
the grand canonical ensemble more commonly used in adsorption
simulations, in order to facilitate the calculation of the heat
capacity.  Two Markov Chain Monte Carlo simulation methods were
used, the Metropolis algorithm~\cite{metropolis,frenkel-smit} and
the Wang-Landau algorithm.\cite{wang-landau,wang-landau2,landau-tsai-exler}
The Metropolis algorithm was used to calculate configurational
observables, such as density distributions and correlation functions.
The Wang-Landau algorithm was used to calculate thermodynamic
observables expressible in terms of ensemble averages or their
derivatives, such as specific and isosteric heat, for certain $N$;
the Metropolis algorithm was used to determine the full $N$ dependence.

The Metropolis algorithm proposes new configurations and accepts
them with a probability equal to
$\min{\{1,P(\textbf{x}^\prime)/P(\textbf{x})\}}$, where $P(\textbf{x})$
and $P(\textbf{x}^\prime)$ are the probabilities of the old and new
configurations $\textbf{x}$ and $\textbf{x}^\prime$; this acceptance
rule causes the random walk to converge to the probability
distribution $P(\textbf{x})$.  By choosing $P(\textbf{x})$ proportional
to the Boltzmann factor $\exp{[-\beta U(\textbf{x})]}$, where
$U(\textbf{x})$ is the potential energy of the configuration, the
Metropolis algorithm uniformly samples configuration space.
For the Metropolis simulations of each $(N,T)$, $4\times10^7$ Monte
Carlo moves were discarded during the initial equilibration to
converge the algorithm to the Boltzmann distribution, then $4\times10^6$
moves were generated, from which $10^4$ samples were drawn to perform
simulated measurements of system observables.

The Wang-Landau algorithm, like the Metropolis algorithm, also
proposes and accepts configurations with a probability equal to
$\min{\{1,P(\textbf{x}^\prime)/P(\textbf{x})\}}$.  However, it chooses
$P(\textbf{x})$ to be proportional to $P[U(\textbf{x})] =
1/g[U(\textbf{x})]$, where $g(U)$ is the (relative) density of
states, thus uniformly sampling {\em energy} space (instead of
configuration space, as in the Metropolis algorithm).  It dynamically
refines its estimate of the density of states by counting each visit
to a state of a given energy $U$ (or, rather, within a small range
of energies $U \in [U_i-\epsilon/2,U_i+\epsilon/2]$ about an energy
bin $U_i$ of width $\epsilon$), and multiplies its running estimate
of $g(U_i)$ by a constant factor $f$.  It continues the random walk
until each energy is visited approximately uniformly (a ``flat
histogram'' of visits in energy space), whereupon it reduces the
factor $f\rightarrow f^{1/2}$ and starts another iteration.  The
algorithm terminates when $f$ is reduced to a preset minimum greater
than unity, with values closer to unity yielding more accurate
estimates of the density of states.

Once an estimate of $g(U)$ is produced, it can be used to calculate
the partition function directly
\begin{equation}
Z \sim \int d\textbf{x}\,e^{-\beta U(\textbf{x})} \approx \sum_i g(U_i) e^{-\beta U_i}\,.
\end{equation}
Thermodynamic quantities can then be calculated from the partition
function, as usual.  One advantage of the Wang-Landau algorithm
over the Metropolis algorithm (and the main reason for using it for
this study) is that, because temperature dependence appears only
in the Boltzmann weight $\exp(-\beta U)$ and not in the density of
states $g(U)$ itself, a single simulation of $g(U)$ can calculate
thermodynamic observables for all temperatures at once.

Some modifications and improvements to the original published
Wang-Landau algorithm were implemented.  Boundary effects were
properly handled.\cite{schulz-etal}  To adapt the original
lattice-based algorithm to continuum systems, preliminary Metropolis
runs at low temperature were performed to estimate a lower bound
on the energy bins (i.e., the ground state energy).\cite{shell-etal}
The simulation can also become trapped for long periods of time in
regions of high degeneracy, so that energies with small $g(U)$ go
a long time before being revisited.  To remedy this, the energy
bins can be broken up into overlapping subranges; ergodicity can
be achieved more rapidly if the interval of energies to be traversed
is smaller.  Separate simulations are performed in each subrange,
producing independent estimates of $g(U)$.  Some care must be taken
in combining them into an estimate of $g(U)$ over the full energy
range:  because each simulation calculates only the {\em relative}
density of states, the estimates will not generally match up at the
boundaries of the subranges.  To overcome this, each subrange
estimate of $g(U)$ is rescaled by a constant factor that minimizes
the least-square error in $\log{g(U)}$ wherever two neighboring
subranges overlap in energy.\cite{shell-etal}  This corresponds
to choosing normalizing factors $C_n$ that minimize the sum $\sum_i
\{\log{[g_n(U_i)/C_n]}-\log{[g_{n-1}(U_i)]}\}^2$ over the overlapping
bins $U_i$ (where $g_n$ denotes the density of states simulated
over subrange $n$), and then rescaling $g_n(U)$ by $C_n$.

For the Wang-Landau simulations of each $N$, 1500 equal-sized energy
bins were used in a range [$U_\textrm{min}$,0], where $U_\textrm{min}$
is the ground state energy.  The 1500 bins were divided into four
overlapping subranges, simulated separately, consisting of the bins
numbered 1--150, 76--787, 713--1425, and 1351--1500.  A histogram was
considered ``flat'' when the number of visits to any particular
energy bin was less than $\pm 20$\% the average number of visits
to any bin.  The minimum $f$ factor was $f_\textrm{min} = 1+10^{-5}$.

\section{Correlation functions}
\label{sec:dens-corr}

\begin{figure}[htb]
\includegraphics[width=8.5cm]{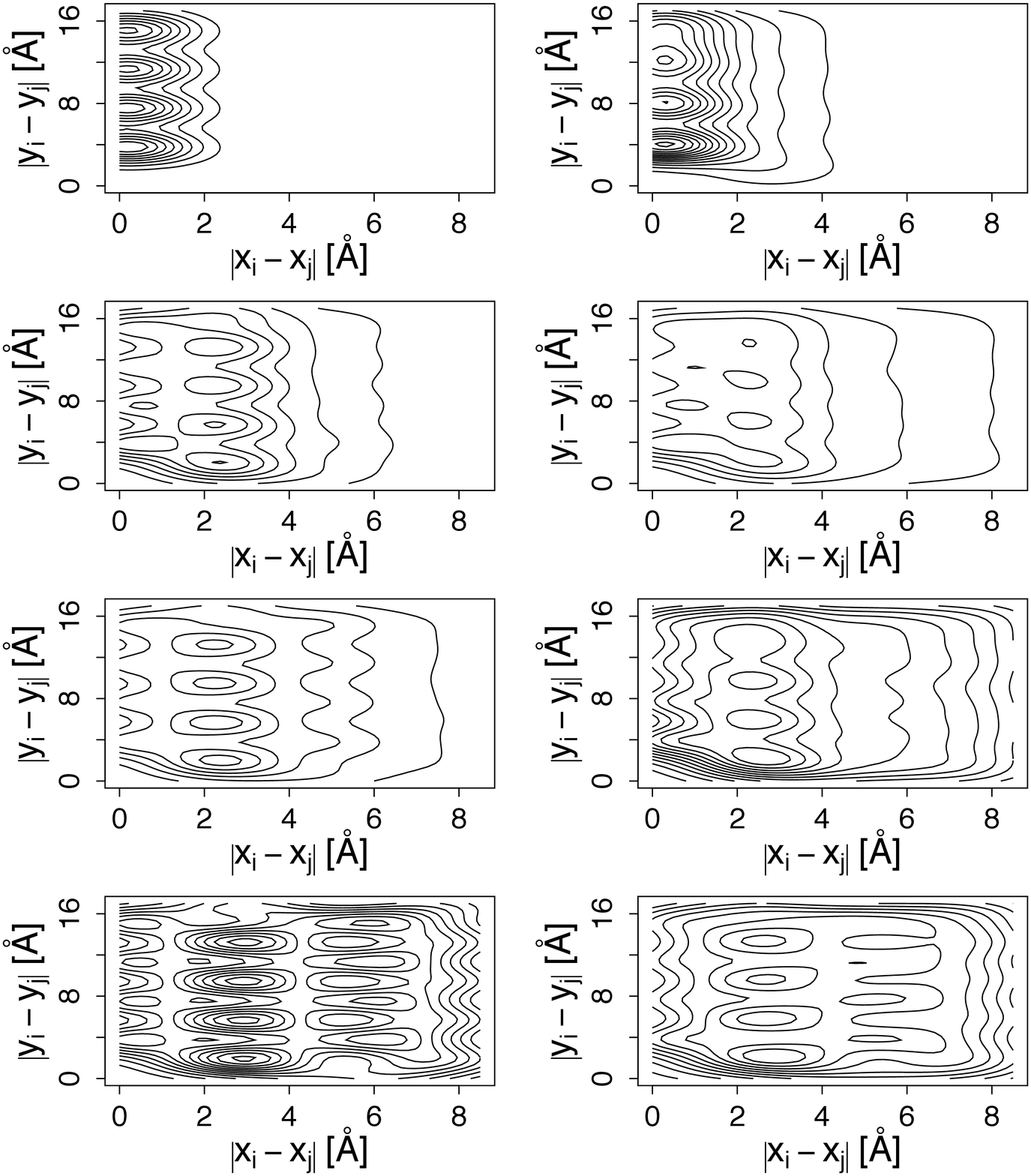}
\caption{\label{fig:corr-xy}
Equiprobability contour plots of the projection into the $xy$ plane
of the pair correlation function, $G(x,y)$ for, from top to bottom,
$N=9$ (groove), $N=18$, $N=27$ (three-stripe), and $N=54$ (monolayer),
at $T=60$~K (left) and $T=90$~K (right).  Distances are in
angstroms.  Note that the vertical axis scale is compressed
relative to the horizontal.}
\end{figure}

For the purposes of this paper, the three-dimensional pair correlation
function is defined as the probability density $g(\mathbf{r})$ that
two particles are separated by a relative displacement $\mathbf{r}$.
Its projection $G(x,y) \equiv \int dz\,g(\mathbf{r})$ into the $xy$
plane is depicted in Fig.~\ref{fig:corr-xy}.  The contours become
wider and more irregular at higher temperatures, as the particles
are thermally excited out of their well-defined low temperature
sites.

In the top pair of panels, one observes the nearly periodic,
quasi-one-dimensional (1D) order within the groove.  As studied
recently in connection to nanotube
adsorption,\cite{carraro1,carraro2,fisher,radhakrishnan-gubbins,wilson-vilches,gatica-calbi-cole}
this phase may undergo a phase transition due to the weak interactions
between particles in neighboring grooves.  In the second pair of
panels, one observes that the correlations within the three-stripe
phase are weaker and even more $T$ dependent than those in the
groove phase.  At 90~K, at higher coverage (seen in the middle four
panels), the stripes are not as straight, primarily due to transverse
excitation (as discussed in Sec.~\ref{sec:heat-cap:higher-coverage}).
Note that the half-filled stripe case ($N=18$) is somewhat less
ordered than the completely filled three-stripe case ($N=27$), as
is expected.  The bottom panels of Fig.~\ref{fig:corr-xy} exhibit
a highly correlated anisotropic two-dimensional (2D) solid at 60~K,
the order of which washes out nearly completely by 90~K, as the
monolayer melts.

Previous experimental studies of Ar adsorption onto planar
graphite~\cite{shaw-fain,larher,migone-li-chan,zhang-larese} found
that the melting temperature depends on density, starting near 55~K
at low density, and increasing with density.  Although the system
studied here differs from that experiment in geometry, we expect
the melting temperature of the monolayer in our system to similarly
increase with density.  The corrugation of the nanotube bundle
should elevate the melting temperature somewhat compared to a planar
surface, as the grooves will serve to more strongly confine the
film's structure.

Bienfait~\textit{et al}.\ have measured diffraction patterns for Ar
on nanotube bundles.\cite{bienfait-etal}  Probably due to heterogeneity
(for which there exists evidence in bare surface diffraction), the
diffraction data are not easily interpreted.  There is, however,
definite evidence of 1D interatomic spacing (i.e., a peak at wave
vector $q=17/\textrm{nm}$) at low coverage and 2D close-packed
spacing (peak near $20/\textrm{nm}$).

\section{Heat capacity}
\label{sec:heat-cap}

\subsection{Overview}
\label{sec:heat-cap:overview}

The isochoric specific heat, $c(T) = (\partial{E}/\partial{T})_V/N$,
i.e., the heat capacity (per particle) at constant volume, was calculated
from ensemble averages.  It is known~\cite{pathria} that the
heat capacity can be given in terms of energy fluctuations,
$(\partial{E}/\partial{T})_V = (\mean{E^2}-\mean{E}^2)/(k_B T^2)$,
where $\mean{\cdot}$ denotes an expectation taken over the canonical
ensemble.  The equipartition theorem gives the kinetic energy
contribution of $\frac{1}{2}k_B$ per degree of freedom to the
specific heat, yielding a total $c(T) = \frac{3}{2}k_B +
(\mean{U^2}-\mean{U}^2)/(N k_B T^2)$.  Given the density of states
$g(U)$ calculated with the Wang-Landau algorithm, the expectations
may be calculated from $\mean{U} = Z^{-1} \sum_i U_i\,g(E_i)
\exp{(-\beta U_i)}$, and similarly for $\mean{U^2}$.  The heat
capacity was also estimated directly from the derivative
$(\partial{E}/\partial{T})_V$ by means of a finite difference
approximation.  These latter estimates, while consistent with the
fluctuation estimates, were ``noisier'' and are not considered
further in this paper.

\begin{figure}[htb]
\includegraphics[height=8.5cm,angle=-90]{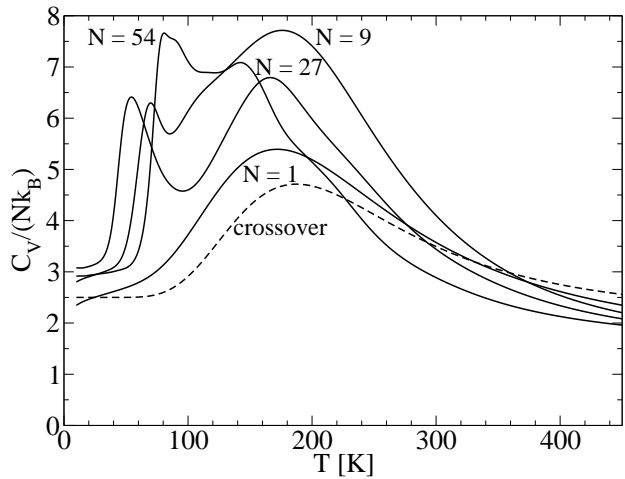}
\caption{\label{fig:cvt}
The dimensionless specific heat $c(T)/k_B$ of the low-density limit
($N=1$); the groove ($N=9$), three-stripe ($N=27$), and monolayer
($N=54$) phases; and the theoretical prediction of the low-density
limit given by the dimensional crossover model (discussed in the
text).}
\end{figure}

\begin{figure}[htb]
\includegraphics[width=8.5cm]{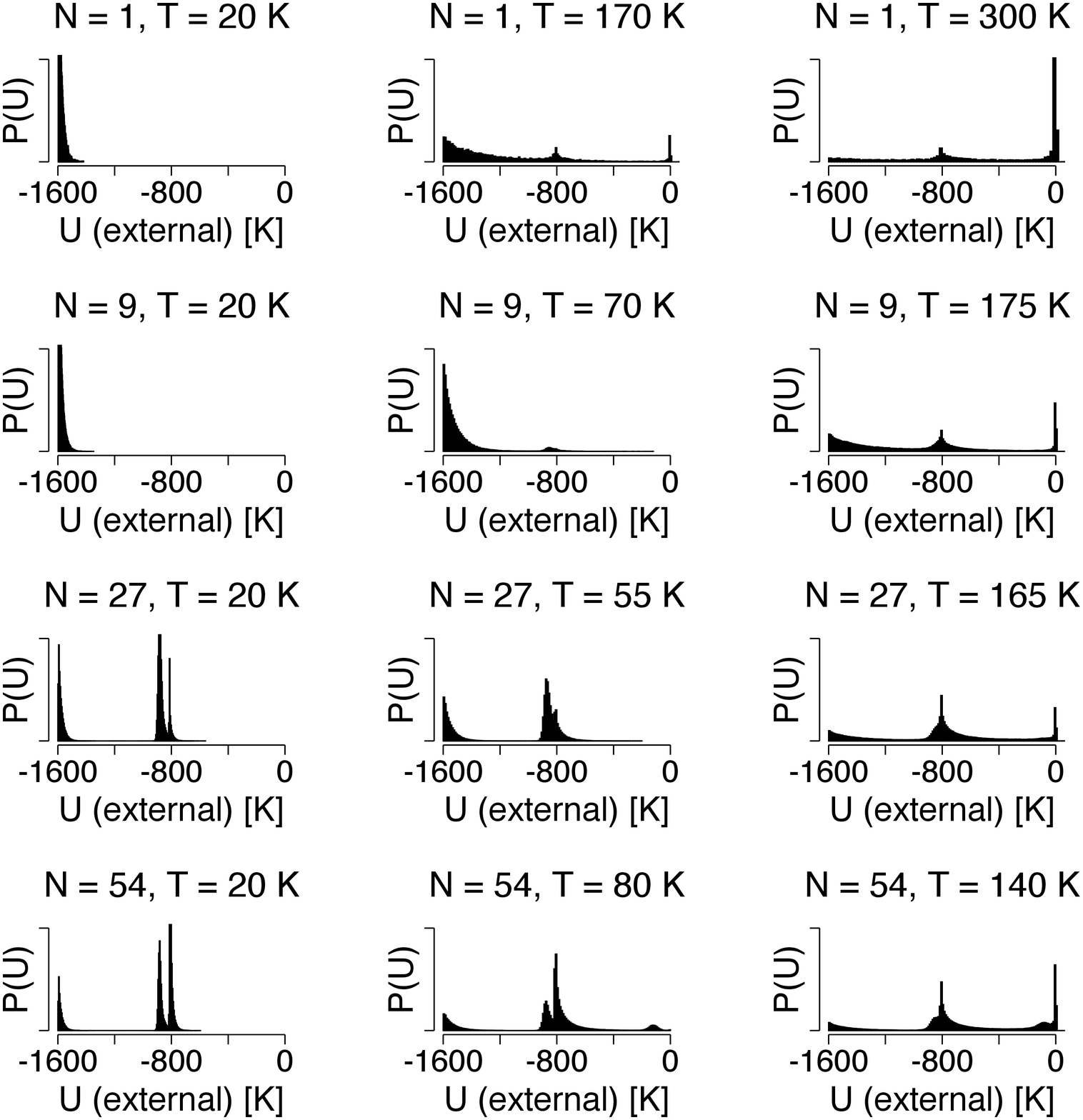}
\caption{\label{fig:uext}
The relative probability $P(U_\textrm{ext})$ of finding a particle
at an external potential of $U_\textrm{ext}$, for the low-density limit
($N=1$), and the groove ($N=9$), three-stripe ($N=27$), and monolayer
($N=54$) phases, for various temperatures.}
\end{figure}

The simulated $c(T)$ for the groove, three-stripe, and monolayer
phases is shown in Fig.~\ref{fig:cvt}.  Note that the overall trend
is for $c(T)$ to have a remarkably high value, in the range
3--7~Boltzmanns, much higher than might be expected from simple
quasi-one-dimensional and two-dimensional models.  We do not have
a detailed quantitative model to explain all of the observed features,
but we can give a qualitative explanation of its behavior.  The
explanations are justified by examining the probabilities of finding
particles at given energies in the external potential, Fig.~\ref{fig:uext},
indicating the fraction of particles that are in the groove,
stripes/monolayer, etc.\@ (quantified in Table~\ref{tab:uext}).

\begin{table}
\begin{tabular}{c|c||c|c|c}
$N$ & $T$~(K) & \% groove & \% monolayer & \% vapor \\
\hline
  & 20 & 100 & 0 & 0 \\
1 & 170 & 61 & 31 & 7 \\
  & 300 & 14 & 35 & 51 \\
\hline
  & 20 & 100 & 0 & 0 \\
9 & 70 & 93 & 6 & 0 \\
  & 175 & 43 & 42 & 15 \\
\hline
   & 20 & 33 & 67 & 0 \\
27 & 55 & 33 & 67 & 0 \\
   & 165 & 22 & 60 & 18 \\
\hline
   & 20 & 17 & 83 & 0 \\
54 & 80 & 17 & 77 & 7 \\
   & 140 & 14 & 54 & 32
\end{tabular}
\caption{\label{tab:uext}
Percentage of particles in the groove
($-1600\textrm{ K}<U_\textrm{ext}<-1200\textrm{ K}$), monolayer
($-1200\textrm{ K}<U_\textrm{ext}<-400 \textrm{ K}$), and vapor
($U_\textrm{ext}>-400\textrm{ K}$) regimes, for the low-density
limit ($N=1$), and the groove ($N=9$), three-stripe ($N=27$), and
monolayer ($N=54$) phases, for various temperatures.}
\end{table}

\subsection{Low density}
\label{sec:heat-cap:low-density}

Consider first the low-density limit.\cite{calbi-cole}  At low
temperatures, the specific heat is near 2.5~Boltzmanns.  This is to be
expected:  the three kinetic degrees of freedom each contribute the
usual $1/2$~Boltzmann; the two transverse dimensions, for which the
external potential is approximately harmonic at its minimum at the
center of the groove, each contribute another $1/2$~Boltzmann.  As
the temperature increases, a peak in the specific heat occurs near
170~K when substantial numbers of adatoms are promoted out of the
groove and into monolayer sites elsewhere on the surface of the
nanotubes (see Table~\ref{tab:uext}).  As $T\rightarrow\infty$, the
adatoms desorb from the surface altogether, and $c(T)$ approaches
the 3/2~Boltzmanns of the three kinetic degrees of freedom of a
pure vapor.  (This will be the case for all other densities as well,
in the high-$T$ limit.)

\begin{figure}[htb]
\includegraphics[height=8.5cm,angle=-90]{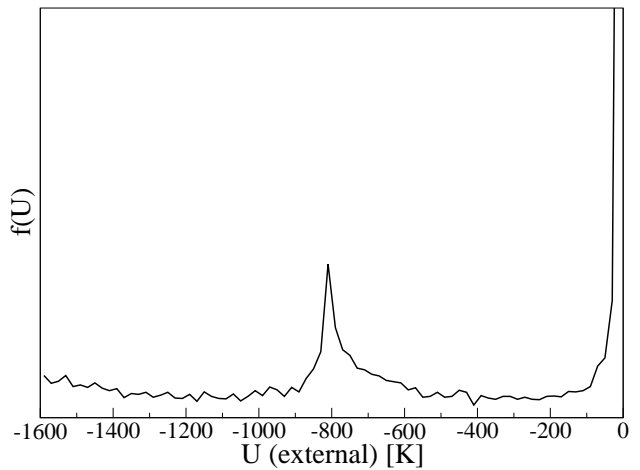}
\caption{\label{fig:vol-dos}
The function $f(U)$ (unscaled), giving the volume of space $f(U)\,dU$
enclosed within a range of external potential energy $[U,U+dU]$.}
\end{figure}

These conclusions are corroborated, as mentioned, in the first row
of Fig.~\ref{fig:uext}; these results, in the low-density limit,
can also be understood by examining the so-called ``volume density
of states''~\cite{trasca-etal} $f(U)$, defined such that $f(U)\,dU$
is the volume of space bounded by infinitesimally separated
isopotential surfaces, $U < U_\textrm{ext} < U+dU$.  This function
is related to the energy probability density $P(U)$ in Fig.~\ref{fig:uext}
at low density by $P(U) = \rho f(U) \exp{(-\beta U)}$, where $\rho$
is the number density of particles.  By dividing an estimate of
$P(U)$ at a given temperature (here, $T=300$~K) by the exponential
Boltzmann factor, we obtain an estimate proportional to the volume
density of states $f(U)$, depicted in Fig.~\ref{fig:vol-dos}.

The qualitative form of this figure can be explained by appealing
to a previously studied analytic model, the dimensional crossover
model.\cite{calbi-cole}  This exactly soluble model ignores
interparticle interactions (an assumption appropriate for the
low-density limit) and treats the nanotube bundle as consisting
solely of two regions, a one-dimensional groove region approximated
by a harmonic potential in the two transverse dimensions, and a
two-dimensional planar monolayer region approximated by a harmonic
potential normal to the surface.  Evaporation from the monolayer
to vapor is neglected.  Its (configurational) partition function is given by
\begin{multline}
Z_\textrm{crossover} = \int_\textrm{groove} d^2 r\,e^{-\beta [V_g + (1/2)\alpha r^2]} \\
+ L_s \int_\textrm{mono} dz\,e^{-\beta [V_m + (1/2)k_m z^2]}\,,
\end{multline}
where the parameters $V_g = -1671$~K, $\alpha = 4898$~K/{\AA}$^2$,
$V_m = -853$~K, and $k_m = 4792$~K/\AA$^2$ were determined by
a fit to the external potential, $L_s = 18$~{\AA} is the approximate
width of the monolayer region in the transverse direction, and
integrations were taken over regions extending 2~{\AA} away from
the groove minimum and 1~{\AA} away from the monolayer minimum.

For the low energies dominated by the groove phase, the crossover
model approximates the external potential as $U = V_g + \frac{1}{2}\alpha
r^2$.  The cylindrical volume enclosed by an isopotential goes like
$V\sim r^2$, and $f(U) = dV/dU = (dV/dr)/(dU/dr)$, which is a
constant; indeed, the $f(U)$ calculated in Fig.~\ref{fig:vol-dos}
is nearly constant at low energies.  For the monolayer region, close
to the potential minimum, $U = V_m + \frac{1}{2}k_m z^2$.  The
rectangular volume enclosed by an isopotential goes like $V\sim z$,
and $f(U) = (dV/dz)/(dU/dz)$, which goes like $z^{-1} \sim
(U-V_m)^{-1/2}$ for $U>V_m$.  This divergence in the model accounts
qualitatively for the peak in $f(U)$ just above the monolayer energy
of about $-800$~K.  For high energies dominated by the vapor phase,
we can treat the substrate as a semi-infinite rectangular volume,
and approximate the external potential by a long-distance ($r^{-6}$)
Lennard-Jones potential integrated over this region, which yields
$U \sim -z^{-3}$.  Then $f(U)$ will go like $z^{-4} \sim (-U)^{-4/3}$,
qualitatively accounting for the sharp rise in $f(U)$ as $U\rightarrow
0$.

Given the volume density of states $f(U)$, we can then obtain the
energy probability density $P(U)$ at any other temperature by scaling
this temperature-independent function by the appropriate Boltzmann
weight.  In particular, the three columns of the first row of
Fig.~\ref{fig:uext} are just the function in Fig.~\ref{fig:vol-dos}
scaled by Boltzmann factors $\exp{(-\beta U)}$ that decay with
decreasing rapidity as $T$ increases ($\beta$ decreases).  The
groove is highly populated at low temperatures (large $\beta$) when
the exponential damping is great enough to suppress population at
higher energies.  The monolayer becomes populated at intermediate
temperatures when the damping is no longer sufficient to suppress
the peak in $f(U)$ that occurs at the monolayer energy ($-800$~K),
and the vapor becomes populated at still higher temperatures (small
$\beta$) when the damping fails to suppress the rapid increase in
$f(U)$ towards 0~K.

The partition function of the crossover model can also be used to
calculate the specific heat directly.  As seen in Fig.~\ref{fig:cvt},
the correspondence between the prediction of this analytic approximate
model and the simulated full model is quite good.

\subsection{Higher coverage}
\label{sec:heat-cap:higher-coverage}

Next, consider the groove phase.  At low temperatures, the specific
heat is near 3~Boltzmanns.  A contribution of 2.5~Boltzmanns is
accounted for by the same argument as for the low density limit.
Unlike the low density limit, however, the groove phase is densely
packed with adatoms, and interparticle interactions must be considered.
An additional $1/2$~Boltzmann arises from confinement in the
longitudinal dimension, for which the Ar-Ar interaction potential
is approximately harmonic at its minimum when the adatoms are stably
distributed in equilibrium.  As the temperature increases, evaporation
out of the groove begins.  At $T\approx 70$~K, evaporation is great
enough to excite adatoms out of the groove; while not many of these
atoms reach the monolayer region (Table~\ref{tab:uext}), there is
still a large change in potential energy for a small increase in
temperature, and thus a large specific heat.  The specific heat
then decreases slightly with increasing temperature, since the
change in energy is not as large once the initial adatoms have begun
to be promoted.  This low-temperature peak is not present in the
low-density case because, as seen in both Table~\ref{tab:uext} and
Fig.~\ref{fig:corr-xy}, the adatoms are not spread transversely as
greatly about the immediate groove region at low temperatures in
the low-density case as they are in the groove case.  However, in
a manner qualitatively analogous to the low-density limit, an
additional, larger peak in the specific heat is found at still
higher temperatures ($T\approx 175$~K), mostly from promotion from
the groove into the stripes and the rest of the monolayer.

Like the groove phase, the three-stripe phase starts out at 3~Boltzmanns
at low temperatures, similar to the groove phase, except that the
1~Boltzmann from external potential confinement in the transverse
plane is replaced by 1/2~Boltzmann from external potential confinement
normal to the surface, and 1/2~Boltzmann from interparticle confinement
along the surface in the transverse plane.  The specific heat peaks
near $T=55$~K; this is not due to a significant fraction of particles
being promoted from the groove to the stripes, as one might expect,
but rather to a wider range of energies within the stripe/monolayer
region, and promotion from the stripes to the rest of the monolayer;
see Fig.~\ref{fig:uext} and Table~\ref{tab:uext}.  It peaks again
near $T=165$~K, with as the groove empties into the stripes and
monolayer, as well as the beginning of evaporation off the surface.

The monolayer phase also starts out at 3~Boltzmanns at low temperatures,
for reasons analogous to the three-stripe phase.  At higher temperatures,
there is a peak in the specific heat near $T=80$~K which, like the
three-stripe peak, is largely due to a broadening of the particles
across a range of energies in the monolayer region, as well as some
promotion from the monolayer to the bilayer (Fig.~\ref{fig:uext}).
Another peak appears near $T=140$~K, corresponding to evaporation
out of the monolayer into the bilayer, and to vapor.

\subsection{Further results}
\label{sec:heat-cap:further-results}

\begin{figure}[htb]
\includegraphics[height=8.5cm,angle=-90]{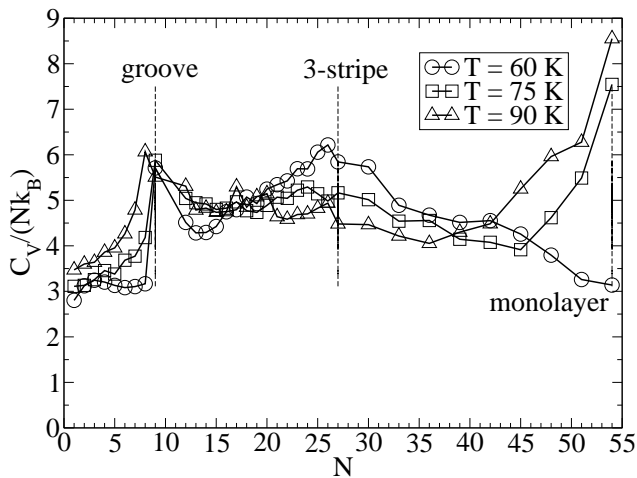}
\caption{\label{fig:cvn}
The dimensionless specific heat $c(T)/k_B$ as a function of density,
at $T=60$, 75, and 90~K.}
\end{figure}

The $N$ dependence of several isotherms is displayed in Fig.~\ref{fig:cvn}.
Of particular note is the rapid rise in the specific heat near
$N\sim 8$, just before groove completion, as $T$ goes from 60 to
90~K.  This is attributed to promotion out of the groove.  Similarly,
near monolayer completion the marked increase in $c$ with $T$ is
attributed to thermal promotion out of the monolayer.

It is also illuminating to study the differential heat of adsorption,
$q_d(N) = -(\partial{E}/\partial{N})_T$, the energy required to adsorb
an additional particle onto the surface at constant temperature.
The differential heat is related to the heat capacity at constant
density, $C_N = (\partial{E}/\partial{T})_N$, by a Maxwell relation
derived from the total derivative $dE = (\partial{E}/\partial{T})_N\,dT
+ (\partial{E}/\partial{N})_T\,dN$, which yields
\begin{equation}
\left (\frac{\partial{C_N}}{\partial{N}}\right )_T = -\left (\frac{\partial{q_d}}{\partial{T}}\right )_N\,.
\label{eq:c-q}
\end{equation}
(Note that $C_N=C_V$ in the canonical ensemble.)

\begin{figure}[htb]
\includegraphics[height=8.5cm,angle=-90]{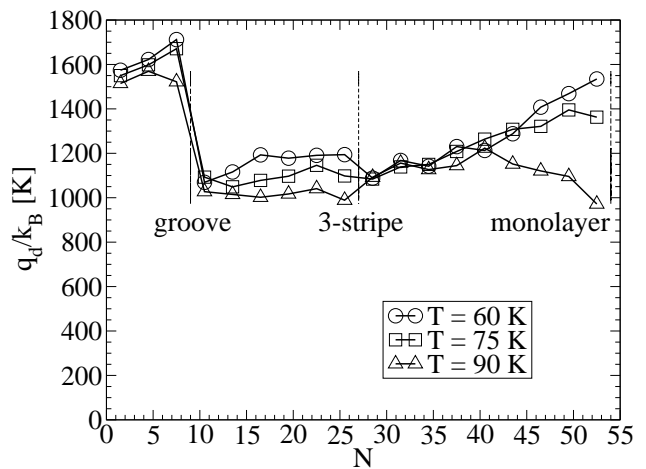}
\caption{\label{fig:q}
The differential heat of adsorption $q_d(N)/k_B$ at $T=60$, 75, and 90~K.
[Shift each curve upwards by its temperature $T$ to obtain the isosteric
heat $q_{st}(N)/k_B$.]}
\end{figure}

The differential heat of adsorption is summarized in Fig.~\ref{fig:q}.
At low densities, the differential heat is near the minimum of the
external groove potential $\approx -1600$~K, becoming slightly
larger at lower temperatures.  Both this value and the $T$ dependence
at low $N$ can be understood from the low-density equation of state,
$U/N = V_g + \frac{5}{2}k_B T$.  As additional particles are added,
the energy for each additional particle is reduced by slightly more
than this amount, to include the interaction energy.  As the groove
phase is approached, the groove becomes tightly packed and the
interaction energy becomes significant, so that adding an additional
particle reduces the energy by the external groove potential plus
the Lennard-Jones well depth, $\epsilon\approx -120$~K for Ar.  The
difference in $q_d(N)$ between low and high temperatures is
particularly large just before the groove phase, which in accordance
with Eq.~(\ref{eq:c-q}) corresponds to the steepest increase in heat
capacity as seen in Fig.~\ref{fig:cvn}; at the $N=9$ groove phase
itself, where the heat capacity peaks with increasing $N$, we see
correspondingly little difference in the differential heat at various
temperatures.  At the other extreme, near monolayer completion, a
similar $T$ dependence is observed. The large decrease in $q_d$
with increasing $T$ is consistent with Eq.~(\ref{eq:c-q}) and
Fig.~\ref{fig:cvn}; the latter shows a large value of $dC/dN$ except
below 60~K.  The explanation is monolayer-to-bilayer promotion above
60~K.

Experimental measurements of isosteric heat for argon on nanotube
bundles have been reported by Wilson \textit{et al}.,\cite{wilson-etal}
Talapatra, Rawat, and Migone,\cite{talapatra-rawat-migone} Jakubek
and Simard,\cite{jakubek-simard} and Bienfait \textit{et
al}.;\cite{bienfait-etal} grand canonical Monte Carlo simulations
have been published by Shi and Johnson.\cite{polydisperse}

The isosteric heat calculations of Shi and Johnson for adsorption
of Ar on a homogeneous bundle at 90~K agree closely with our results,
with a peak of $q_{st}=14$~kJ/mol just before the groove phase,
corresponding to our peak of 1650~K.  Past the groove phase, their
calculated isosteric heat drops and remains constant with coverage,
slightly below 10~kJ/mol, corresponding to our nearly constant value
near 1200~K.

Shi and Johnson compared their calculations to the experimental
results of Wilson \textit{et~al}.\ and Talapatra \textit{et~al}.,
and since we agree with those calculations, we will briefly summarize
their conclusions.  Our calculations agree with both experiments
at higher coverage, beginning at the three-stripe phase, but their
isosteric heats at lower coverage are dramatically greater than
ours, as large as 18~kJ/mol ($\sim 2200$~K) at low coverage.  We
ascribe this discrepancy with experiment to our neglect of bundle
heterogeneity, following Shi and Johnson, whose simulations of
heterogeneous bundles agreed well with both experiments.

In contrast, the isosteric heats measured by Jakubek and Simard
agreed well with our simulations, with a peak of 137~meV ($\sim
1600$~K) near the groove, descending to plateau of 106~meV ($\sim
1200$~K) through to monolayer coverage.  This agreement with our
calculations suggests that their bundles were more homogeneous than
those studied in the other two experiments.  It should be noted,
however, that their isosteric heat continues to decrease as coverage
increases, whereas our isosteric heat appears to rise slightly as
the monolayer is approached.  The isosteric heat of Wilson
\textit{et~al}.\ also drops past the monolayer.

Like the other experiments, the results of Bienfait \textit{et~al}.\
for Ar exhibit two plateaus in the dependence of the isosteric heat
on coverage.  The lowest coverage data yield $q_{st}=15$~kJ/mol,
or about 1800~K.  Our predicted value in this range is of order
1650~K.  The higher coverage, broad plateau corresponds to a measured
$q_{st}=1200$~K, which agrees well with the value we find for the
three-stripe phase.  However, the data at monolayer coverage continue
to decrease, while ours appear to increase, as noted.  Another area
of disagreement is the extent, in coverage, of these plateaus.  In
the data, the second plateau extends over a coverage range comparable
to that of the first plateau.  Our calculations, instead, find that
groove region of high $q_{st}$ extends over just one-sixth of the
range of the combined three-stripe plus monolayer regime (grouped
together because of similar values of $q_{st}$).  This discrepancy
may be attributed to the role of large interstitial cavities within
the bundle, as argued by Bienfait~\textit{et~al}.

\section{Summary and conclusions}
\label{sec:summary}

Our results are intended to stimulate further experimental studies
of this system and analogous systems involving other gases on
nanotube bundles.  We have investigated the variation of thermodynamic
properties with $T$ and $N$.  One of the more interesting general
results is that the specific heat is typically larger than might
have been expected from either simple models used to treat these
systems (either independent particles or a solid)~\cite{siber1} or
from experimental results for films on graphite.\cite{dash-etal}
For most conditions studied here, the specific heat exceeds three
Boltzmanns, with average values in the range four to five Boltzmanns.
In contrast, the specific heat of independent particles~\cite{calbi-cole}
in this environment is less than three Boltzmanns, except at high
$T$ (above 100~K), when the particles are excited out of the groove.
The large values found in these simulations arise from the fact
that the highly corrugated potential surface presents a sequence
of excitation steps
(groove$\rightarrow$three-stripe$\rightarrow$monolayer$\rightarrow\cdots\rightarrow
$vapor), each of which enhances the specific heat.

The temperature dependence of the specific heat shows a characteristic
double-peak structure.  All densities show a large peak near 175~K,
corresponding to promotion of adatoms out of the groove into the
monolayer region.  The groove, three-stripe, and monolayer phases show
an additional peak at lower temperature, corresponding to a thermal
broadening in the range of external potential energies of the
particles, rather than to any significant promotion of particles
into qualitatively different regions.

Other principal results involve the relation between the evolution
of film structure (with increasing $N$) and the corresponding
thermodynamic and correlation functions. As the groove begins to
fill ($N$ approaching 9), the heat capacity shows a dramatic jump
as a function of coverage. Consistent with the Maxwell relation,
Eq.~(\ref{eq:c-q}), the differential heat decreases with $T$ at
that point (Fig.~\ref{fig:q}).  Analogous behavior occurs near
completion of the three-stripe phase, near $N=27$.

Our study has been fully classical, but the temperatures beneath
which quantum effects become significant can be
estimated.\cite{calbi-cole}  We estimate that quantum effects can
be ignored above about 80~K (see Appendix~\ref{sec:qm}).  This is
a higher temperature than some of the important structure in the
heat capacity---the first peak in the heat capacity occurs at or
below this temperature.  Modifications to the heat capacity from
quantum mechanics at very low energies are given by Debye
theory:\cite{siber1,siber2,siber3,kostov-calbi-cole}  we expect
that $c(T)\rightarrow 0$ as $T\rightarrow 0$, and $c(T) \propto
T^d$, where $d\sim 1$ for the groove and $d\sim 2$ for the monolayer,
if the density is high enough to form a bulk phase.  To evaluate
quantum effects accurately would require application of the path
integral Monte Carlo method to the problem.\cite{wang-johnson}

We note, also, that an experimental heat capacity cell has a volume
on the order of 1--10~cm$^3$, whereas our simulation volume was on
the order of $10^{-20}$~cm$^3$.  Our simulation, focusing on small
volume nearer the adsorbed film, thus ignores almost all of the
volume in which desorption into vapor can occur.  This causes the
simulation to underestimate the heat capacity that will be
experimentally measured.  The effects of desorption cannot be ignored
when the number of atoms in the vapor starts to approach the number
of atoms in the film; this occurs at roughly 25--50~K (see
Appendix~\ref{sec:desorption}).

Particularly interesting results from the correlation function
studies include the reduced longitudinal correlations in the groove
and striped phases as $T$ rises above 60~K.  These results would
be amenable to testing by diffraction experiments even if the samples
included a randomly oriented batch of nanotubes; this is a familiar
problem dealt with in powder averaging of small-sample experiments.

This paper studied a system of identical nanotubes.  The sensitivity
of $c(T)$ to nanotube heterogeneity, with an asymmetric groove
region between nanotubes of different sizes, is a potentially
interesting subject for future investigation.\cite{polydisperse}

\begin{acknowledgments}
We are very grateful to David Goodstein for a helpful explanation
of the thermodynamics of adsorption, to Oscar Vilches for a discussion
of experimental issues, to Mary~J. Bojan for discussions of the
simulation methods and their interpretation, and to Michel Bienfait
for his helpful comments.  This work was supported by the National
Science Foundation.
\end{acknowledgments}

\appendix

\section{Quantum effects}
\label{sec:qm}

The upper bound on the temperature at which quantum effects must
be considered is dominated by the physics of the deepest energy
well, i.e., the groove.  We can obtain one estimate by considering
the minimum energy of longitudinal phonons in the groove in Debye
theory, $\hbar \omega_D$, where $\omega_D = \sqrt{k/m}$ and $k =
2^{8/3}(9\epsilon/\sigma^2)$ is the force constant of a quadratic
approximation to the minimum of the Ar-Ar interaction potential, a
12-6 Lennard-Jones potential $U_\textrm{int} = 4\epsilon
[(\sigma/r)^{12}-(\sigma/r)^6]$, with $\sigma=3.4$~{\AA},
$\epsilon=120$~K for Ar.  The corresponding energy is 27~K.  This
estimate considers only Ar-Ar interactions and ignores the external
potential; we can obtain a complementary estimate by ignoring the
interactions and considering only the external potential.  We return
to the crossover model of the groove, outlined in
Sec.~\ref{sec:heat-cap}, as a two-dimensional harmonic oscillator
with a force constant $\alpha = 4898$~K/{\AA}$^2$.  Treating it now
as a \textit{quantum} harmonic oscillator, it is excited at an
energy $\hbar \omega_\perp$, where $\omega_\perp = \sqrt{\alpha/m}$
and $m$ is the atomic mass of argon.  The corresponding energy of
this second estimate is 77~K.  Taking the larger of the two as a
conservative estimate, we expect that quantum effects can be ignored
above about 80~K.

\section{Effects of desorption}
\label{sec:desorption}

We can estimate the temperature at which desorption into the full
volume of an experimental cell becomes significant, by determining
when the ratio $N_v/N_m$ of atoms in the vapor to atoms in the
monolayer becomes significant.  Extending the crossover model, we
can consider the monolayer and vapor phases as separate systems,
the monolayer modeled as a surface with a harmonic normal potential,
and the vapor modeled as a free gas.  The ratio $N_v/N_m$ is then
given by the ratio of their respective partition functions
\begin{equation}
\frac{N_v}{N_m} = {\int_\textrm{cell} dz} \Big{/} {\int_\textrm{mono} dz\,e^{-\beta [V_m + (1/2)k_m z^2]}}\,.
\end{equation}
We take the first integral between zero and the cell height, $h$;
the second integral may be safely taken between zero and $\infty$,
in the interests of finding an analytic solution.  This gives
\begin{equation}
\frac{N_v}{N_m} = \frac{h}{\sqrt{\pi/(2\beta k_m)}}\,e^{-\beta V_m}\,.
\label{eq:vap-mono}
\end{equation}
The fraction of atoms in the vapor above which desorption should
be considered ``significant'' is ambiguous, but we might take it
to be 10\%--20\%.  Solving Eq.~(\ref{eq:vap-mono}) for $\beta$,
using $h=1$~cm and the values for $V_m$ and $k_m$ found in
Sec.~\ref{sec:heat-cap:low-density}, this corresponds to a temperature
in the range of 25--50~K.

This calculation neglects interparticle interactions.  Their inclusion
would lower the estimate of the temperature at which desorption
from the monolayer into vapor becomes significant, analogously to
how the evaporation from the groove to the monolayer takes place
at a lower temperature when the groove is packed---the interacting
case---than when it is sparsely populated and the adatoms are
effectively noninteracting.  This is supported by the data in
Table~\ref{tab:uext}:  more groove$\rightarrow$monolayer promotion
occurs for $N=9$ at $T=175$~K than for $N=1$ at the comparable
temperature $T=170$~K, indicating that the groove promotion begins
at a lower temperatures for the interacting $N=9$ than for the
noninteracting $N=1$.

\end{document}